\documentclass[prl,twocolumn,superscriptaddress,nofootinbib]{revtex4-2}

\usepackage{lipsum}
\usepackage{braket}
\usepackage{graphicx}
\usepackage{dcolumn}
\usepackage{enumitem}
\usepackage{bm}
\usepackage{amsmath}
\usepackage{mathtools}
\usepackage{subfigure}
\usepackage{color}
\usepackage{xcolor}
\usepackage{verbatim}
\usepackage{float}
\usepackage{mathrsfs}
\usepackage{natbib}

\newcommand{\ua}[0]{\ket{\uparrow}}
\newcommand{\da}[0]{\ket{\downarrow}}
\newcommand{\exc}[0]{\ket{e}}
\newcommand{\whf}{\omega_{\uparrow\downarrow}}

\begin{document}

\title{Site-dependent selection of atoms for homogeneous atom-cavity coupling}
\author{Baochen Wu$^1$, Graham P.~Greve$^1$, Chengyi Luo$^1$, James K.~Thompson}

\affiliation{JILA, NIST, and Department of Physics, University of Colorado, 440 UCB, 
Boulder, CO  80309, USA}
\date{\today}

\begin{abstract}
    We demonstrate a method to obtain homogeneous atom-cavity coupling by selecting and keeping $^{87}$Rb atoms that are near maximally coupled to the cavity's standing-wave mode. We select atoms by imposing an AC Stark shift on the ground state hyperfine microwave transition frequency with light injected into the cavity. 
    We then induce a spin flip with microwaves that are resonant for atoms that are near maximally coupled to the cavity mode of interest, after which, we use radiation pressure forces to remove from the cavity all the atoms in the initial spin state. 
    Achieving greater homogeneity in the atom-cavity coupling will potentially enhance entanglement generation, intracavity driving of atomic transitions, cavity-optomechanics, and quantum simulations. This approach can easily be extended to other atomic species with microwave or optical transitions. 
\end{abstract}

\maketitle

Atomic ensembles in optical cavities have become a versatile and powerful platform for creating atomic entanglement \cite{hosten2016Nat, cox2016PRL, Norcia2018Sci, Davis2019PRL, Hacker2019NatPhoton}, generating superradiant lasers \cite{Norcia2016SciA,bohnet2012Nat}, synthesizing quantum matter \cite{Clark2020Nature}, interacting many-body pseudo-spin and related systems and precision measurement \cite{Juan2020arxiv,PP2020Nat}. The widely-used cavity geometry consisting of just two mirrors is simple, but the standing wave nature of the modes can lead to inhomogeneity in the coupling of the individual atoms to the cavity.  The inhomogeneous coupling reduces the effective atom number \cite{Hu2015PRA}, creates dephasing in driven systems \cite{Muniz2020Nature}, and brings about optomechanically-induced oscillations \cite{cox2016PRL}, all of which can be undesirable in certain contexts.

In many experiments, one wishes to freeze out the motional degrees of freedom of the atoms by tightly trapping the atoms in a lattice inside the cavity at wavelength $\lambda_l$ that is far-detuned from any atomic resonance. However, the lasers or cavity modes for probing or mediating spin-spin interactions have a standing wave with periodicity determined by a wavelength $\lambda_p$ close to an atomic transition  $\lambda_a$.  There have been previous efforts to circumvent this issue of inhomogeneous coupling, such as using a commensurate trapping wavelength such as $\lambda_l = 2\lambda_p$ \cite{hosten2016Nat, Lee:14}, However, the wavelength of the trapping laser could be a degree of freedom we want to reserve for tuning, for example,  to engineer state-dependent traps and magic-wavelength traps \cite{Katori2009PRL, Ludlow2006PRL, Heinz2020PRL}. Tightly confining magnetic traps have also allowed atoms to be loaded into single lattice sites, requiring a high atomic density at fixed total atom number.  If localizing the atoms is undesired, one can also allow the atoms to move along the cavity axis to time-average away the standing-wave inhomogeneity \cite{Cox2016PRA}.  Finally, one can use ring-cavity geometries with running waves \cite{Schine2016Nat}, but this adds complexity, results in larger mode volumes, and breaks the degeneracy between polarization modes of the cavity.


Here we demonstrate a method to obtain more homogeneous atom-cavity coupling by initially loading atoms into thousands of lattice sites and then using a spectroscopic method to select and keep only those atoms at lattice sites with near maximal coupling to the probe cavity mode.  While this approach involves discarding atoms, it has favorable scaling in that the degree of homogeneity is expected to scale quadratically with the number of retained atoms.


\begin{figure}[!htb]
    \centering
    \includegraphics[width=\columnwidth]{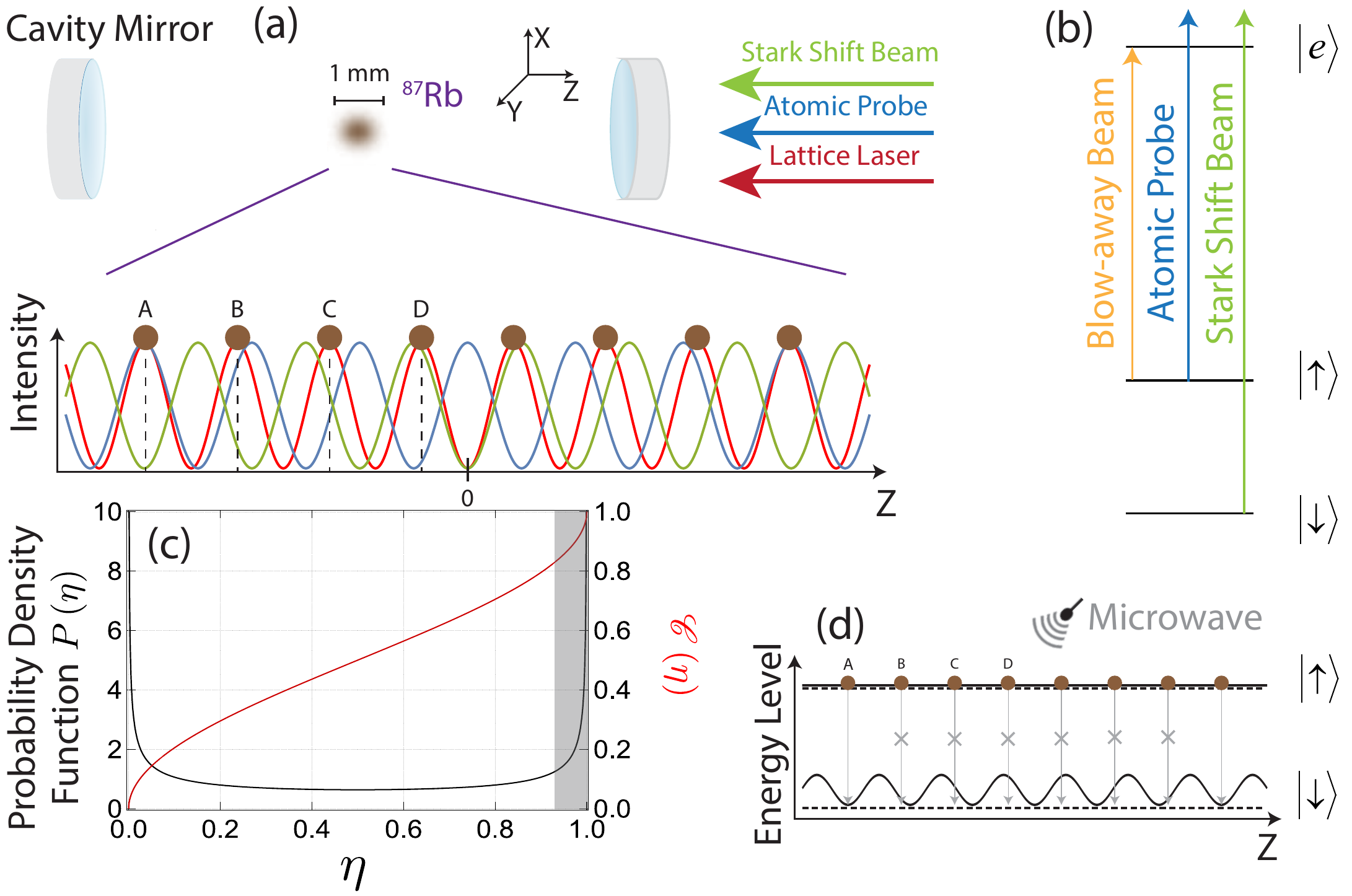}
    \caption{ Experimental setup and working principle. (a) Three lasers are sent into the cavity for trapping (lattice laser), probing (atomic probe) and dressing (Stark shift beam) the atoms. The atoms are always trapped at the maxima of the optical lattice. The intensity of the atomic probe and the Stark shift beam the atoms see are site dependent and opposite near the center of the cavity. (b) Level diagram of $^{87}Rb$, and the relationship between the lasers and energy levels. (c) Atomic probability distribution $P(\eta)$ as a function of $\eta = g^2/g_0^2$ for atoms trapped in the optical lattice without selection. The gray region indicates the atoms we want to select. The cumulative distribution function $\mathscr{C}(\eta)=\int_0^1 P(\eta) d\eta$ is also presented to confirm that $P(\eta)$ is normalized. (d) Principles of the selection. The dashed lines are the original energy levels without the presence of the Stark shift beam. The solid lines are the Stark-shifted energy levels. Atoms with unchanged microwave transition frequencies are maximally coupled to the atomic probe, and can be selected with microwave $\pi$-pulses at well tuned frequencies.} 
    \label{fig:expt_setup}
\end{figure}


To further explain the problem and our approach, let us consider our experimental system in which initially $N \approx 10^5$ $^{87}$Rb atoms are trapped at the center of the cavity using an intracavity optical lattice at $\lambda_L =813~$nm. The atoms are trapped at antinodes of the lattice light as shown in Fig.~\ref{fig:expt_setup}a red. Two additional lasers, the atomic probe and Stark shift beams, can be injected into the cavity at or near resonance with other longitudinal modes of the cavity.  Their functions in the spectroscopic selection process will be returned to shortly.  

A simplified level diagram is shown in Fig.~\ref{fig:expt_setup}b.  The two relevant hyperfine ground states here are the magnetic-field insensitive states $\ua\equiv\ket{F=2, m_F=0}$ and $\da\equiv\ket{F=1, m_F=0}$ with a splitting of 6.834~GHz.  Only the optical excited state $\exc$ denoting $\ket{F'=3, m_F=0}$ is shown for simplicity with transition wavelengths $\lambda_a \approx 780$~nm. 

Here, we wish to enhance the homogeneity of the coupling of the atoms to the atomic probe.  The atomic probe and an associated longitudinual cavity mode is typically detuned by $\Delta_p/2\pi=700$~MHz from resonance with the $\ua$ to $\exc$ transition.  The atomic-probe can then be used to measure the shift in the frequency of the cavity mode by an amount $\delta_c \approx \left(g_0^2/\Delta_p\right)\Sigma_i^N\cos^2\left( 2 \pi z_i/\lambda_p\right)$ \cite{Chen2014PRA}, where $2 g_0$ is the single-particle vacuum Rabi frequency, atom $i$'s position along cavity axis relative to the cavity mid-plane is $z_i$, and only atoms in $\ua$ are implicitly included in the sum.  This same frequency shift has been used previously to realize entanglement generation via collective measurements, cavity-optomechanics, and cavity-mediated spin-spin interactions \cite{hosten2016Nat, cox2016PRL, Norcia2018Sci, BrooksNature2012}.  The goal is then to select atoms at antinodes of the probe mode (\textit{i.e.}~$\cos^2\left (2 \pi z_i/\lambda_p\right)\approx 1$), and remove all other atoms from the cavity mode.  To reduce notational complexity, we define the coupling $g_i$ of atom $i$ relative to that of a maximally coupled atom as $\eta_i \equiv \left(g_i/g_0\right)^2$.

In conceptual analogy to what is done in NMR imaging by applying magnetic field gradients \cite{PurcellNMR, BlochNMR}, here we engineer a spatially-dependent shift of the $\da$ to $\ua$ transition frequency $\whf \rightarrow \whf + \delta\left(z\right)$.  The frequency shift is generated by an AC Stark shift $\delta\left(z_i\right) = \delta_{s}\sin^2\left( 2 \pi z_i\lambda_{s}\right)$ that is induced by injecting light into a cavity mode one free spectral range away from that of the probe mode.  Adjacent longitudinal modes have the desired opposite symmetry.  The free spectral range of the cavity 6.791 GHz is close to the hyperfine splitting $\whf$, so that the Stark shift mode is detuned by approximately $\Delta_s/2\pi \approx 700$~MHz from the $\da$ to $\ket{e}$ transition, leading primarily to a Stark shift of the state $\da$.  Because of the small frequency difference between the probe mode and the Stark shift mode, the standing-waves are to good approximation locally out of phase near the center of the cavity where the atoms are located.

To select atoms with peak coupling to the probe mode, we begin by optically pumping atoms into $\ua$, applying the spatially-dependent Stark shift beam, and then applying microwaves for $t_m\approx 200$ to $250 ~\mu$s at frequency $\omega_m =  \whf +\delta_m$ to perform a $\pi$-pulse for atoms whose transition frequency is resonant with the microwaves.  The bandwidth of transition frequencies that undergo the spin flip is set by the Rabi frequency $\Omega_m$ of the applied microwaves. Atoms that are not flipped to $\da$ are then removed from the trap using a radiation pressure force, which is transverse to the cavity axis, from the blow away beam that is tuned to resonance with the $\ua$ to $\ket{e}$ transition and applied for 100~$\mu$s.  If the microwaves are resonant with the unshifted transition frequency, \textit{i.e.} $\delta_m=0$, then atoms at nodes of the Stark beam will remain in the trap. We depicted the selection process in Fig.~\ref{fig:expt_setup}d, where atom A will survive after the blow-away stage, while atoms B, C and D will be removed. In principle, the remaining atoms can be optically pumped back to $\ua$ and the selection process can be repeated for improved performance. We also plot the probability density of atoms as a function of $\eta$ in Fig.~\ref{fig:expt_setup}c. The grey region indicates the selection bandwidth set by $\Omega_m$.

\begin{figure}[ht]
    \centering
    \includegraphics[width=\columnwidth]{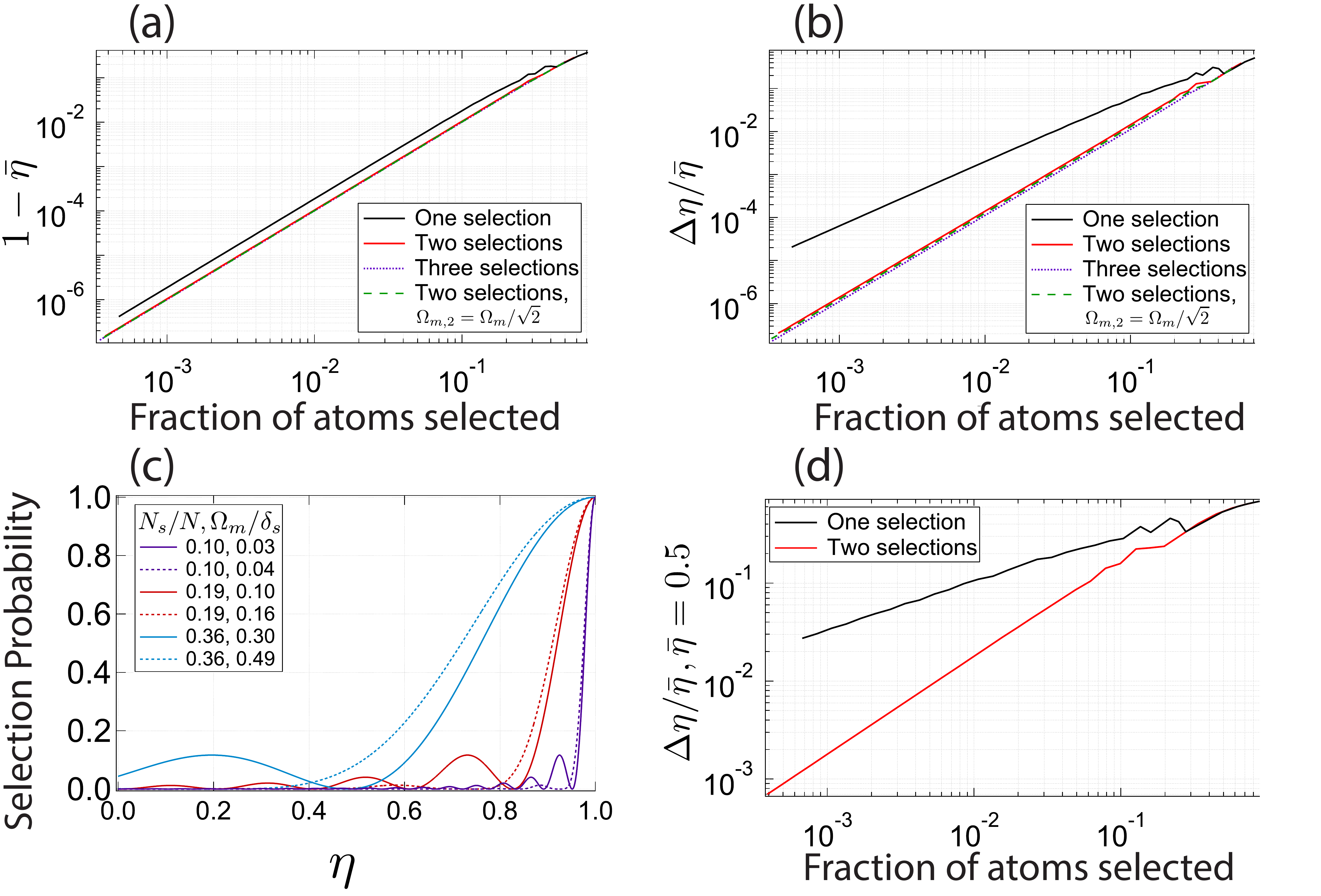}
    \caption{Simulation with minimal model. (a) Curves for $1-\bar{\eta}$ vs fraction of atoms selected, for one, two and three selections, coded in black solid, red solid and purple dashed. We also plotted the same relationship for two selections with varying Rabi frequencies, with the Rabi frequency of the second selection pulse $\Omega_{m,2} = \Omega_m\sqrt{2}$, which is coded in green dashed. The curve for three selections is hard to identify because it is pretty much buried under the red and green curves. The simulation is done with $\delta_m=0$, and the change in the fraction of atoms is achieved by changing the ratio $\Omega_m/\delta_s$. (b) Curves for $\Delta\eta/\bar{\eta}$ vs fraction of atoms selected, for one, two and three selections, coded in the same way as in (a). The curve for two selections with varying Rabi frequencies is also presented. (c) Transfer probability vs $\eta$ for one selection (solid curves) and two selections (dashed curves). Different colors correspond to different $\Omega_m / \delta_s$ and $N_s/N$. (d) Curves for $\Delta\eta/\bar{\eta}$ vs fraction of atoms selected, for one and two selections, color coded in black and red. These curves are obtained with $\delta_m=\frac{1}{2}\delta_s$.
    }
    \label{fig:theory}
\end{figure}

The probability that an atom with a given coupling is selected is simply given by the Rabi spin flip probability as

\begin{equation}
\begin{array}{l}
F\left(\eta_c, \eta, \Omega_m, \delta_s\right)=\\
~~~~~~~~~~\frac{\Omega_m^2}{\Omega_m^2+\left(\left(\eta_c-\eta\right) \delta_{s}\right)^2} \sin^2\left(\frac{\pi}{2} \frac{\sqrt{\Omega_m^2+\left((\eta_c-\eta) \delta_{s}\right)^2}}{\Omega_m}\right),
\end{array}
\label{eqn:WinFcn}
\end{equation}

\noindent where the microwave detuning is parameterized as $\eta_c = \delta_m/\delta_s$. 
In Fig.~\ref{fig:theory}c we show the spin flip probability for a given coupling $\eta_i$ for different ratios of $\Omega_m/\delta_s$ at $\eta_c=0$.

%
In Fig.~\ref{fig:theory}a, we show the theoretical trade-off between the fraction of of atoms retained after selection $N_s/N$ and the ensemble averaged coupling $\bar{\eta}\equiv\left<\eta_i\right>$ for one, two, three and four selections. In Fig.~\ref{fig:theory}b, we show the trade-off versus the fractional standard deviation of the coupling about the mean $\Delta\eta/\bar{\eta}\equiv \sqrt{\left<\eta^2\right>-\bar{\eta}^2}/\bar{\eta}$.  In this model, the microwave detuning is $\delta_m=0$ and the atomic ensemble's axial spatial extent is much larger than the differential wavelength $\lambda_d^{-1} = \left|\lambda_l^{-1}-\lambda_p^{-1}\right|$.  From numerical simulations such as in Fig.~\ref{fig:theory}a and \ref{fig:theory}b, we find in the region $N_s/N <0.1$ that  $1-\bar{\eta}= A \left(N_s/N\right)^\alpha$, where $\alpha = 2.00$ and $A= 1.83, 1.04, 1.00, 1.00$ for 1, 2, 3 and 4 selection pulses with the same Rabi frequency respectively. The fitting result for two selections with varying Rabi frequencies is $\alpha=2.00$ and $A=1.01$. Similarly, we find  $\Delta\eta/\bar{\eta} = A \left(N_s/N\right)^\alpha$, where $\alpha = 1.47, 2.00, 2.00, 2.00$ and $A= 1.75, 1.41, 1.12, 1.09$ for 1, 2, 3, and 4 selection pulses with the same Rabi frequency respectively. The fitting result for two selections with varying Rabi frequencies is $\alpha=2.00$ and $A=1.27$. For two selections or greater, the qudratic scaling of the inhomogeneity is highly favorable.  We also note that the inhomogeneity is only marginally improved for greater than two selection pulses. Applying two selection pulses with varying Rabi frequencies can improve the averaged coupling and homogeneity compared to two selection pulses with the same Rabi frequency, but it cannot beat three selection pulses.

For comparison, we show the trade-off versus the fractional standard deviation of the coupling about the mean $\Delta\eta/\bar{\eta}$ when $\delta_m=\frac{1}{2}\delta_s$ in Fig.~\ref{fig:theory}d. We could select atoms located at the maximum slope of the atomic probe intensity profile in this way, which are potentially useful in certain circumstances, for example, ponderomotive squeezing \cite{BrooksNature2012}. Because of symmetry, $\bar{\eta}=0.5$ holds all the time for these atoms, as noted in the figure. Likewise, we find $\Delta\eta/\bar{\eta}=A\left(N_s/N\right)^\alpha$, where $\alpha = 0.48, 0.99, 1.00$ and $A= 0.93, 1.70, 1.30$ for 1, 2 and 3 selection pulses with the same Rabi frequency respectively. Only linear scaling can be obtained with two or more selection pulses.

\begin{figure}[ht]
    \centering
    \includegraphics[width=\columnwidth]{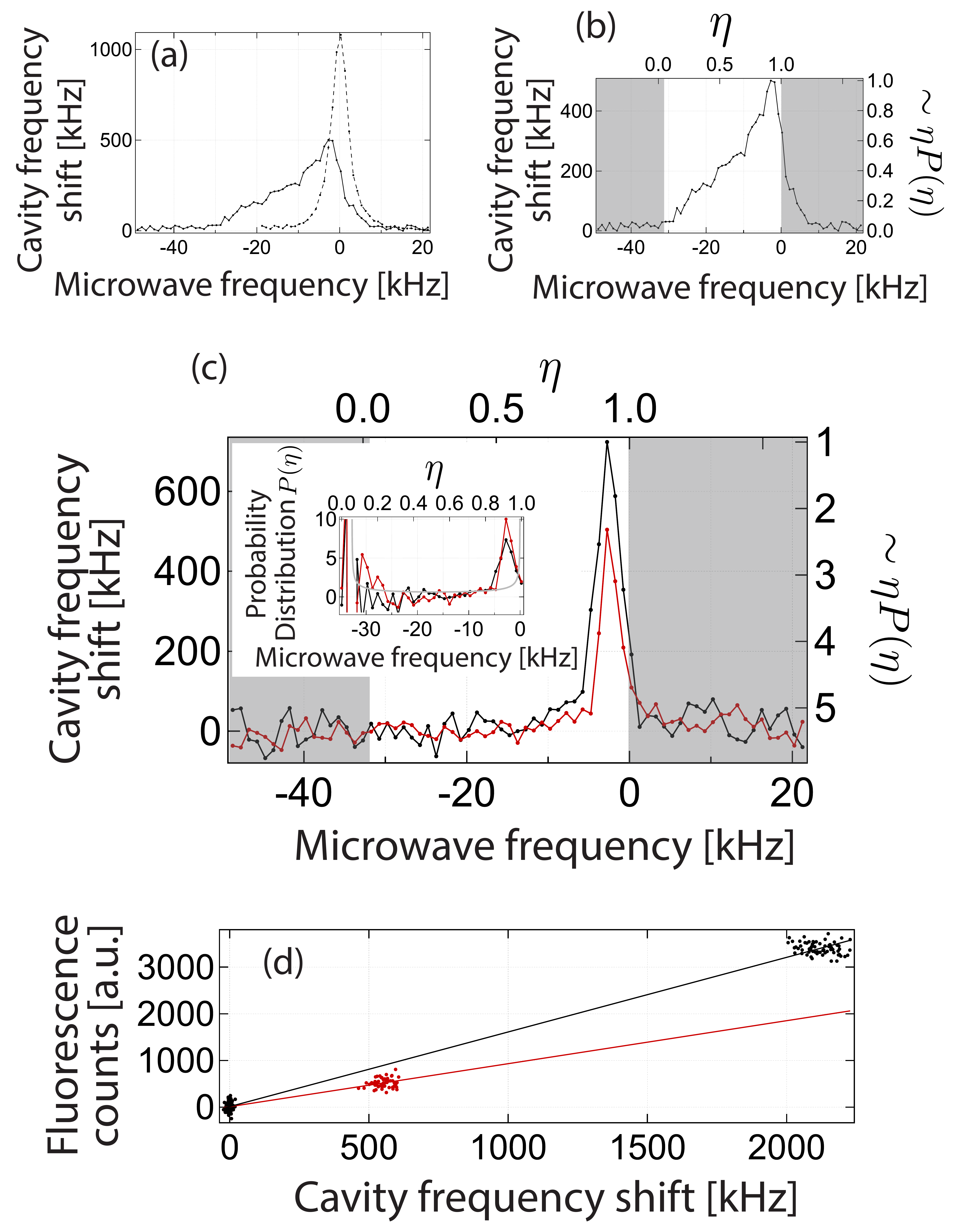}
    \caption{Microwave spectrums and fluorescence measurements. (a) Microwave spectrums for atoms with (solid line) and without (dashed line) the Stark shift beam. For reference, a single atom sitting at the anti-nodes of the atomic probe gives a cavity frequency shift of about 150 Hz. (b) Microwave spectrum for atoms with the Stark shift beam on in the cavity frequency shift vs microwave frequency basis can be mapped onto approximate $\eta^2P(\eta)$ vs $\eta$ basis, with $P(\eta)$ the atomic probability distribution. Physically unavailable regions are shaded in gray. (c) Microwave spectrums for atoms after one (black) and two (red) selections are mapped in the same way as the spectrum in (b). The insets are in $P(\eta)$ vs $\eta$ basis obtained after approximating window functions as delta functions. The grey line in the background is the original probability density function for atoms without any selections. (d) Spectrums in fluorescence counts vs cavity frequency shift basis for atoms after one selection (red) and without any selections (black). }
    \label{fig:microwave_spec}
\end{figure}

How the Stark shift beam modified the microwave spectrum is shown in Fig.~\ref{fig:microwave_spec}a. The dashed trace was the original microwave spectrum without the presence of the Stark shift beam with long-time Rabi flopping. The solid trace was taken with microwave $\pi$-pulses of Rabi freqeucny $\Omega_{m0}=2\pi\times2.04$~kHz, when the Stark shift beam was on. The horizontal axis could also be converted into $\eta=g^2/g_0^2$, which the microwave transition frequency change was proportional to, as shown in Fig.~\ref{fig:microwave_spec}b. $\eta$ was physically meaningful in the range from 0 to 1, and the corresponding microwave frequency range suggested that the peak AC Stark shift $\delta_{s0}$ induced by Stark shift beam was about $2\pi\times32.7$~kHz. The finite cavity frequency shift in the grey regions came from non-zero microwave Rabi frequency. The cavity frequency shift $\delta\omega_c$ after a microwave $\pi$-pulse takes the form in Eq.~\ref{eqn:TransProb},


\begin{equation}
    \delta\omega_c=N_{tot} \delta\omega_{c0} \int_0^1\eta P(\eta) F(\eta_c, \eta, \Omega_p, \delta_s)  d\eta,
    \label{eqn:TransProb}
\end{equation}

\noindent where $N_{tot}$ is the total number of atoms, $\delta\omega_{c0}$ is the cavity frequency shift induced by a  single  atom  sitting  at  the  anti-node  of  the  atomic probe, $P(\eta)$ is the probability density, and $F(\eta_c, \eta, \Omega_p, \delta_s)$ is as defined in Eq.~\ref{eqn:WinFcn}, with $\Omega_p$ the Rabi frequency for spectroscopy $\pi$-pulse.

$F(\eta_c)$ will approach Dirac delta function $\delta(\eta_c)$ in the limit $\Omega_p \rightarrow 0$. As a result, the vertical axis of Fig.~\ref{fig:microwave_spec}b could only be approximated as $\eta P(\eta)$ due to finite $\Omega_p$, represented by the tilde on the axis label. It is worth mentioning that we used $\Omega_p = \Omega_{m0}$ for the spectrum in Fig.~\ref{fig:microwave_spec}b, and $\Omega_p \approx 1/12~\Omega_{m0}$ for the spectrums in Fig.~\ref{fig:microwave_spec}c.

The black and red traces in Fig.~\ref{fig:microwave_spec}c were the microwave spectrums for atoms after one selection pulse and two selection pulses. The spectrums were taken with microwave $\pi$-pulses of Rabi frequency $\Omega_p=2\pi\times0.17$~kHz, and $\delta_m=-2\pi\times2.7$~kHz. Likewise, the horizontal and vertical axes could be mapped to $\eta$, and $~\eta P(\eta)$. If we anyway approximated $F(\eta_c)$ as Dirac delta function $\delta(\eta_c)$, we got the discrete probability density distribution $P_{d}(\eta)$, normalized to the total area under the curve, truncated in the range $0.4<\eta<1$, as shown in the inset of Fig.~\ref{fig:microwave_spec}c. We did not include $\eta<0.4$ in calculating the area because the data points in that range were too noisy, since the cavity frequency shift was comparable to our measurement noise floor there. The results are summarized as follows. For one selection, we measured $\bar{\eta}=0.88(3)$, $\Delta\eta/\bar{\eta}=0.14(3)$, $N_s/N = 0.11(1)$, and predicted $\bar{\eta}=0.90$, $\Delta\eta/\bar{\eta}=0.12$, $N_s/N = 0.18$. For two selections,  we measured $\bar{\eta}=0.92(4)$, $\Delta\eta/\bar{\eta}=0.1(1)$, $N_s/N=0.083(8)$, and predicted $\bar{\eta}=0.92$, $\Delta\eta/\bar{\eta}=0.04$, $N_s/N = 0.12$.



Another way to measure $\bar{\eta}$ was by comparing the measured cavity frequency shift to the observed fluorescence from the atoms with and without selections. Atoms contributed differently to the cavity frequency shift depending on their coupling to the cavity, but contributed equally to fluorescence imaging. Fig.~\ref{fig:microwave_spec}d is a plot of fluorescence counts vs cavity frequency shift. The black and red data points were for atoms without selections and after one selection, whose slopes were fitted to be $a_u=1.60(1)$ and $a_s=0.92(3)$ Counts/kHz respectively. We got $\bar{\eta}=a_u/2a_s=0.87(3)$, in agreement with the numerical prediction $\bar{\eta}=0.90$ and the previous estimate from the microwave spectrums $\bar{\eta}=0.88(1)$.

In principle, one would expect that $\bar{\eta}$ could go even closer to 1, if we select atoms with $\delta_m=0$~kHz microwave center frequency. We decided to take the data with detuned microwave pulses for signal to noise reasons.

\begin{figure}[!htb]
    \centering
    \includegraphics[width=\columnwidth]{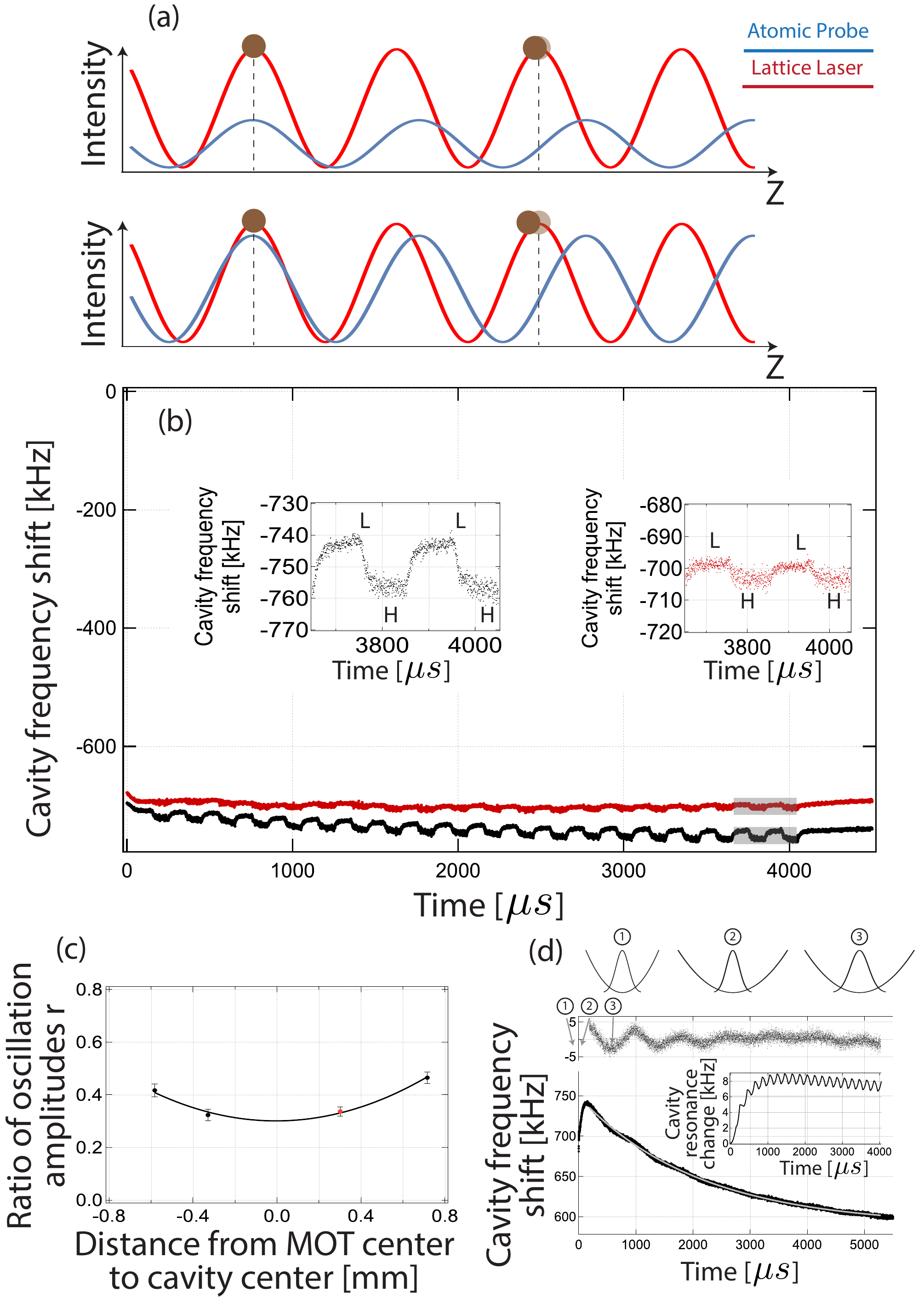}
    \caption{Dressed cavity equilibrium resonance change due to optomechanics. (a) Illustration of how the optomechanical force from the atomic probe could affect the equilibrium position of the atoms, and thus the dressed cavity resonance. The force arises from the gradient of the potential formed by the atomic probe. The partially transparent atoms indicate their original positions. The displacement scales with the atomic probe power. (b) Dressed cavity resonant frequency shift as a function of time when the power of the atomic probe is toggled between high and low power level. The insets are zooming-in of the grey regions. The black and red traces are for atoms without any selections and with two selections. (c) The ratio of fractional amplitudes of oscillations of the cavity resonance frequency with and without selections r, as a function of the distance from the MOT center to the cavity center.  (d) Bottom panel shows the dressed cavity resonance frequency shift as a function of time with a constant atomic probe power, where the grey curve is an exponential fit. The inset is the simulated contribution of the radial oscillations to cavity-resonance frequency in Fig.~\ref{fig:optomec}b as we toggle the power in the atomic probe. The residue of the exponential fit is presented in the middle panel. The top panel sketches the relationship between the trap depth and the distribution of the atomic ensemble in position space, before, at the moment, and after the atomic probe is turned on, as indicated by the grey arrows.}
    \label{fig:optomec}
\end{figure}

Optomechanical-induced oscillations of the cavity resonance, arising from the inhomogeneous atom-cavity coupling, were observed in our previous experiment and were believed to be the main limiting factor \cite{cox2016PRL}. We now characterize the ability of our scheme to suppress this effect. Fig.~\ref{fig:optomec}a explains the cause of the optomechanically-induced oscillations. All atoms are initially trapped at the maxima of the optical lattice. Another potential is developed when we turn on the atomic probe. The gradient of the potential is force, so the equilibrium position of the atoms in the optical lattice will be displaced by an amount that depends on the gradient force induced by the atomic probe, if they are not trapped near the extrema of the atomic probe. The displacement will change the dressed cavity resonance because the atom-cavity coupling is changed. In principle, oscillations of the dressed cavity resonance at the axial trapping frequency are expected whenever the atomic probe power is changed. However, we were only able to see the equilibrium dressed cavity resonance change as we toggled the power of the atomic probe, because the linewidth of our cavity was about 50 kHz, much smaller than the axial trapping frequency of the optical lattice, which was about 205 kHz.

To probe the optomechanical effect, we took a sequence as follows, after two selection pulses, we pumped atoms to $\ket{F=2, m_F=2}$, actively locked the atomic probe frequency to the dressed cavity resonance, and monitored the dressed cavity resonance as we toggled the atomic probe power up and down with windows of 100 $\mu$s long. The polarization of the atomic probe was $\sigma_+$, so that the atom number loss from free space scattering could be ignored on the $\ket{F=2, m_F=2}$ to $\ket{F=3', m_F=3}$ cycling transition. The high and low power levels were different by a factor of 3.3. The results are presented in Fig.~\ref{fig:optomec}b. The insets are zoom-ins of the grey regions, which include two toggling cycles each. The transients of the locks are removed by subtracting off the trace obtained with an empty cavity.


As a figure of merit of the degree of inhomogeneity, we considered the degree to which the optomechanical oscillations were suppressed by the selection process via the ratio $r= a_s/a_0$ of the fractional amplitudes  of oscillations of the cavity resonance frequency with $a_s$ and without $a_0$ selection. We found a measured ratio $r =0.32$ and fraction of retained atoms $N_s/N = 0.04(1)$ with $N=8\times10^4$ after a double selection with $\delta_m=0$ and $\Omega_m/\delta_s=0.08$  in Fig.~\ref{fig:optomec}b.  For comparison, the theoretical predictions were $r=0.08$, $N_s/N = 0.14$ and $\bar{\eta}=0.96$.  To get a sense of the sensitivity of the prediction to system parameters, if we allowed for a small shift in detuning of the applied microwave frequency of $2$~kHz with the same sign as the applied AC stark shift, the simulation predicted $r=0.32$, however this also predicted $N_s/N= 0.14$, and $\bar{\eta}=0.91$ and the source of a possible 2~kHz shift was unknown.  Other effects that had been considered to explain the discrepancy but were not sufficiently large enough included: finite extent of axial wavefunctions, non-AC Stark shift induced inhomogeneous broadening of the microwave transition, finite radial temperature, imperfect $\pi$-pulses, errors in the ratio of the Rabi frequency to peak AC Stark shift $\Omega_m/\delta_s$, finite optical pumping to $F=1$, and potential heating during blow away stage.


We also observed that the toggling of the atomic probe power drove radial oscillations of the atomic ensemble, which made a considerable contribution to the observed oscillation amplitude of the dressed cavity resonance. The atomic probe not only provided an axial force on the atoms, but also changed the radial confining potential. As a result, the finite temperature of the atomic ensemble and a non-adiabatic change in the radial confinement combined to drive a  breathing of the  radial extent of the atomic ensemble, with a period that was half of the radial oscillation period $T_r$ as illustrated in the top panel of Fig.~\ref{fig:optomec}d. Fig.~\ref{fig:optomec}d middle and bottom panels show the observed oscillations of the cavity resonance frequency due to the radial breathing. The oscillations were induced here by turning on the atomic probe and keeping it at a constant power level.  

The inset of Fig.~\ref{fig:optomec}d shows the simulated contribution of the radial oscillations to the cavity-resonance frequency in Fig.~\ref{fig:optomec}b. Applying a correction for the amplitude ratio $r$, we found a corrected $r =0.25$ closer but still off the predicted value of $r=0.08$. This did not, however, affect the discrepancy in the number of selected atoms.


The quality of the selection also depended on the local orthogonality of the standing waves formed by the atomic probe and the Stark shift beam, \textit{i.e.} they were out of phase. We measured $r$ as a function of the position of the atomic cloud along the cavity axis as shown in Fig.~\ref{fig:optomec}c, which was well described by a quadratic fit with a minimum value $r= 0.30(2)$.  
All preceding data presented in this paper was taken at the position indicated by the red point, which was 0.30(4) mm off the cavity center, not sufficiently far enough away to account for any residual discrepancies in $r$ and $N_s/N$.

In conclusion, we have demonstrated a method to obtain homogeneous atom-cavity coupling. We select atoms that are near-maximally coupled to the cavity with the help of Stark shift beam that is one FSR away from the cavity mode we probe the atoms. In this way, we could always keep the atoms tightly trapped (\textit{i.e.} Lamb-Dicke regime) and retain the optical lattice wavelength as a tunable degree of freedom, which opens the route for state-dependent trapping, magic wavelength trapping and other wavelength-dependent tricks. We have also shown that this method helps suppress the optomechanical oscillations, which was a main obstacle for achieving large amount of squeezing in inhomogenous atom-cavity coupling systems \cite{cox2016PRL}. We believe the method demonstrated in this paper could benefit experiments based on atom-cavity interactions in a broad sense.

\begin{acknowledgements}

We would like to thank Chun-Chia Chen and Colin Kennedy for fruitful discussions. This work is supported by NSF JILA-PFC PHY-1734006 grant, NSF QLCI Q-SEnSE 2016244, and NIST.
.

\end{acknowledgements}



\bibliographystyle{apsrev4-2}
\bibliography{Site_Selection.bib}


\end{document}